%% file: main.tex
\pdfoutput=1
\documentclass[acmsmall,nonacm]{acmart} 
\usepackage{url}
\usepackage{amsmath}
\usepackage{wrapfig}

\usepackage{algorithm} 
\usepackage[noend]{algcompatible}
\usepackage{algpseudocode} 

\usepackage{lipsum} 
\usepackage{graphicx}
\usepackage{xcolor}
\newcommand\OA[1]{}
\newcommand\RS[1]{}
\newcommand\DS[1]{}
\usepackage{bm}
\usepackage{caption}
\usepackage{subcaption}
\usepackage{mathpartir}

\usepackage{utf8math}

\usepackage[inline]{enumitem}
\usepackage{float}
\usepackage{ulem}
\usepackage{xcolor}
\DeclareMathOperator*{\argmax}{\textsf{argmax}} 
\DeclareMathOperator*{\similar}{\textsf{sim}} %
\DeclareMathOperator*{\dist}{\textsf{dist}} %
\DeclareMathOperator*{\PC}{\textsf{PC}} %

\usepackage[english]{babel}
\usepackage{amsthm}

\newtheorem*{property}{\itshape Property}

\newtheorem{definition}{Definition}[section] 

\title{Proximal Byzantine Consensus}
\author{Roy Shadmon}
\affiliation{%
  \institution{University of California, Santa Cruz}
  \country{USA}
}
\author{Daniel Spencer}
\affiliation{%
  \institution{Indiana University}
  \country{USA}
}
\author{Owen Arden}
\affiliation{%
  \institution{University of California, Santa Cruz}
  \country{USA}
}

\begin{document}
\normalem 
\input{abstract}

\makeatletter
\let\@authorsaddresses\@empty
\makeatother
\maketitle

\input{introduction}
\input{PC}
\input{system_model}

\input{implementation}
\input{impact}
\input{evaluation}

\input{related}

\input{conclusion}

\bibliographystyle{ACM-Reference-Format}
\bibliography{main} 

\input{appendix}

\end{document}

%% file: abstract.tex
\begin{abstract} 
Distributed control systems require high reliability and availability guarantees, despite
often being deployed at the edge of network infrastructure.  Edge computing resources
are less secure and less reliable than centralized resources in
data centers.  Replication and consensus protocols improve robustness to
network faults and crashed or corrupted nodes, but these volatile 
environments can cause non-faulty nodes to temporarily diverge, increasing
the time needed for replicas to converge on a consensus value, and 
give Byzantine attackers too much influence over the convergence process.

This paper proposes \emph{proximal Byzantine consensus}, a new
\emph{approximate} consensus protocol where clients use statistical models of
streaming computations to decide a consensus value. In addition, it provides an
interval around the decision value and the probability that the true
(non-faulty, noise-free) value falls within this interval.
Proximal consensus (PC)
tolerates unreliable network conditions, Byzantine behavior, and other sources of 
noise that cause honest replica states to diverge.  
We evaluate our approach for scalar values, and compare 
PC simulations against a vector consensus (VC) protocol simulation.
Our simulations demonstrate that consensus values selected by PC
have lower error and are more robust against Byzantine attacks. 
We formally characterize the security guarantees against Byzantine attacks 
and demonstrate attacker influence is bound with high probability.
Additionally, an informal complexity analysis suggests PC
scales better to higher dimensions than convex hull-based protocols such as VC.
\end{abstract}

%% file: introduction.tex
\section{Introduction}

Distributed control systems need timely access to \emph{feedback
data}—data from (or derived from) sensors—in order to make control
decisions.  The consequences of delayed, missing, or corrupted data depends on the
system; from suboptimal performance to catastrophic failure.  Hence,
the lengths a control system designer's willingness to ensure data
integrity and availability is frequently application specific.  A 
manifestation of this dynamic are design parameters that characterize
limits on the noise or
error in the feedback data that the control system can tolerate.  When
a feedback data source processes sensor data in one or more stages
before emitting it to the control system, noise effects accumulate, combining 
the inherent measurement noise of the sensors and noise caused by
missing or delayed inputs at each processing stage.  Feedback data from the
physical world is almost always noisy to some extent, so systems are designed to tolerate an
amount considered reasonable by the system designers.  
For sensitive control system parameters, however,
exceeding the expected threshold could cause unpredictable, even dangerous, behavior.

Tolerating faulty behavior has been a central focus of
distributed systems research for decades. By replicating system components and
requiring a consensus among the components' outputs, fault tolerant
protocols have been developed for many scenarios.  A \emph{crash-fault
tolerant} (CFT) system preserves the safety (a ``bad thing'' never
happens) and liveness (a ``good thing'' eventually happens) of the
system despite up to an upper-bound of $F$ replicas crashing, meaning
they stop responding altogether.  A \emph{byzantine-fault tolerant}
(BFT) system preserves these properties when up to $F$ replicas behave
arbitrarily, such as maliciously deviating from the protocol.

Noisy feedback data presents a challenge for fault-tolerance in
distributed control systems since the noise makes it harder to ensure
non-faulty consensus outputs. At a high level, CFT protocols such as
Paxos~\cite{paxos} and BFT protocols such as PBFT~\cite{pbft} reach
consensus by finding a quorum of replicas proposing identical values,
with the size of the quorum ensuring that (1) a minimum number of
non-faulty hosts are included, and (2) at least one of these
non-faulty hosts participated in the previous phase or
round. Together, these properties ensure that non-faulty nodes 
have a consistent view of the system state, and the system can always
output values corresponding to the outputs of a
quorum of non-faulty nodes (provided no more than $F$ faults occur).
In a distributed control system, however, non-faulty replicas might
propose different values due to various sources of noise.  Thus, a
quorum of identical values may not exist.

Protocols that ensure \emph{approximate} consensus among non-faulty
replicas have been studied to address this challenge.  Many of these
protocols guarantee that consensus values are bound by the \emph{convex
hull} of the non-faulty outputs. The convex hull of a set of points is
the smallest convex shape that contains all the points. In one
dimension, the convex hull is a line interval, in two dimensions it is a
polygon, and so on. The challenge of selecting a value in the convex
hull of non-faulty outputs is that the identity of the faulty replicas
is unknown, so it is not possible to determine the convex hull directly.

\begin{figure}
    \centering   
    \begin{subfigure}{0.49\textwidth}
        \centering
        \includegraphics[width=\linewidth]{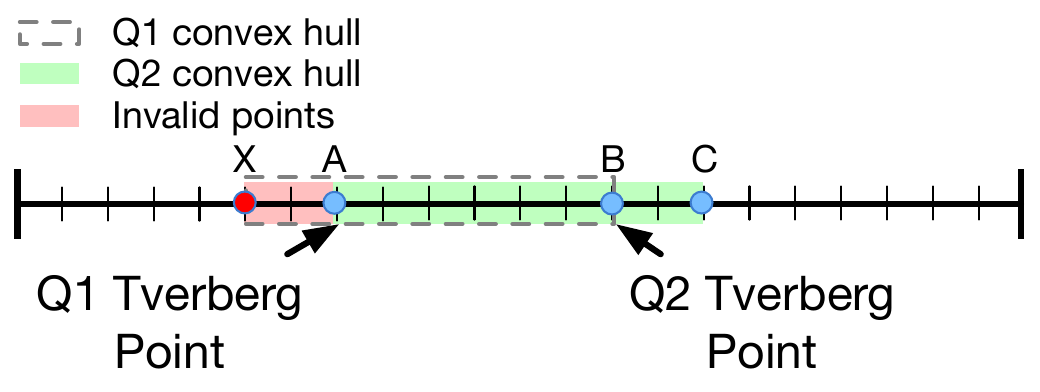}
        \caption{
          Tverberg-based consensus in one dimension. 
        } \label{fig:tverberg}
    \end{subfigure}
    \begin{subfigure}{0.49\textwidth}
        \centering
        \includegraphics[width=\linewidth]{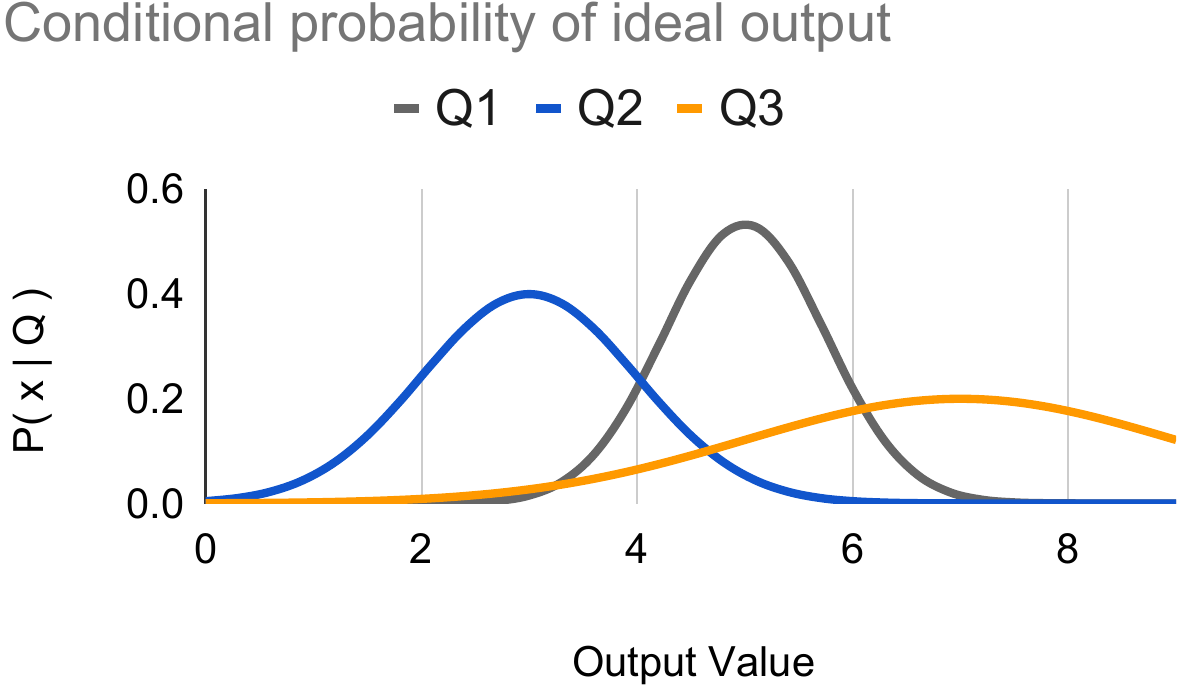}
        \caption{
          Output probabilities from quorum observations.
        } \label{fig:pcgoal}
    \end{subfigure}    
    \caption{Tverberg consensus selects points within the convex hull of non-faulty replicas. 
             Proximal consensus selects points most likely to have been produced by non-faulty replicas. }
\end{figure}

Figure~\ref{fig:tverberg} illustrates a common approach to approximate
consensus based on \emph{Tverberg points}~\cite{byz-vector-consensus}.
Suppose one replica must choose an output
value based on four proposed values: its own value and the values it
received from the three other replicas, one of which may be faulty. In
Figure~\ref{fig:tverberg}, these values are placed on a number line.
The faulty value ($X$) is red and the non-faulty values ($A,B,C$) are blue.
There are four quorums of size $3$, but only two define intervals
containing non-overlapping regions: $Q1=\{X,A,B\}$, and $Q2=\{A,B,C\}$.

By design, all quorums must contain at least two non-faulty values, so
for a particular quorum $Q$, there must be at least one subset $S \subseteq Q$
with $|S|=2$ that contains no faulty values.  The replica then selects
a value in the intersection of the intervals (convex hulls) formed by
each size-two subset.  Tverberg's Theorem~\cite{tverberg} guarantees
this intersection is non-empty.  Notice that even if a network
partition prevented one of the non-faulty values from delivery, the
replica can still select a value in the non-faulty interval of Q1.
When more than one Tverberg point is available, most protocols choose
some aggregate function of the points such as their mean.\footnote{In
one dimension, Tverberg points correspond to the median~\cite{tverberg-median}, but
the idea of convex hulls generalizes to higher dimensions, where
Tverberg points represent a kind of ``higher-dimensional median.'' In
this paper, we focus only on scalar (one-dimensional) consensus.}

The convex hull guarantee\footnote{Other approaches that guarantee
containment in the convex hull of non-faulty outputs use
centerpoints~\cite{vc-centerpoint}, Weighted-Mean Subsequence Reduced
algorithms~\cite{w-msr}, or median-based algorithms~\cite{median-algs} instead of
Tverberg points.  Although we do not discuss these approaches
specifically, much of our analysis regarding consensus values bound by
a non-faulty convex hull apply more generally.} is a strong one:
Byzantine attacks can never cause a consensus value to be outside a
region defined only by non-faulty nodes. However, the larger the convex
hull, the more power an attacker has to influence the output value. 
For example, if the attacker instead chose $X$ in
Figure~\ref{fig:tverberg} to be to the right of $A$, $B$, and $C$, then the
relevant quorum would be \{$B$,$C$,$X$\} and C would be a Tverberg point
instead of $A$. Likewise, if a network partition prevented delivery of
value $C$, any value proposed by the attacker between $A$ and $B$ would be
chosen as the Tverberg point.
In noisy or volatile environments, the distance between non-faulty
outputs is likely to grow, giving attackers more influence over the
chosen value.

This paper presents \emph{proximal byzantine consensus}, an
alternative to convex hull methods for finding consensus values in
noisy environments. Rather than using a geometric basis such as
Tverberg points to select values robustly, a proximal consensus
replica uses statistical inference conditioned on the inputs it
receives from other replicas.  The core insight behind proximal
consensus is that, since they are performing the same deterministic
computation on highly correlated inputs, the outputs of non-faulty
replicas should be similar.  To exploit this similarity, proximal
consensus replicas infer the most probable ``ideal'' output value
(without noise) given the observations it has received.

If all values received by a replica were non-faulty, inferring the
ideal output would be straightforward. Given prior
distributions for the outputs and the noise, we would want to find
the value that maximizes the probability density function (PDF)---the
most probable value---of the unknown output distribution given the
observed values. But since some of the values may be Byzantine,
treating these values as legitimate observations would give attackers
too much influence over the inferred output.  Attempting to model the Byzantine
values statistically is not an option since by definition Byzantine 
replicas behave arbitrarily: an attacker's
previous behavior is not predictive of its future behavior.

Instead, for each possible quorum, proximal consensus infers the
conditional probability distributions \emph{assuming the observed
values were produced by non-faulty nodes}. Figure~\ref{fig:pcgoal}
illustrates three PDFs, each conditioned on a different quorum of
values (Q1, Q2, and Q3) received by a replica. Each curve represents
the probability that the x-axis value is the ideal output if the
values in that quorum were produced by non-faulty nodes.  Since the
conditional probability of an output is proportional to the similarity
of the quorum values, as long as there are a sufficient number of
non-faulty values in each quorum, the quorum that produces the PDF
with the highest maximum value (Q1) is the most likely to contain only
non-faulty values or faulty values that are similar to non-faulty
values. If it contains faulty values, they must be at least as similar
to non-faulty values as other non-faulty values are, otherwise
exchanging those values would increase the maximum of the PDF. Thus,
the replica's best choice based on the quorums in
Figure~\ref{fig:pcgoal} is 5, the most probable output given the
values in Q1.

We discuss below the formal guarantees offered by proximal consensus
and use proximal consensus in the design of two related protocols:
\begin{itemize}
  \item A ``one shot'' protocol where replicas broadcast proposed
    values to clients without requiring coordination and clients locally
    determine the proximal consensus value and confidence interval.
  \item A coordinated protocol where replicas propose values
    until either a desired confidence interval is reached or a minimum
    number of messages are received.
\end{itemize}
These protocols require additional replicas compared to convex hull 
methods: proximal consensus requires a minimum of $4f+1$ replicas compared 
to $3f+1$, but this cost is offset by lower asymptotic complexity
in higher dimensions (Section~\ref{sec:complexity}), stronger
guarantees against Byzantine influence on outputs~(Section~\ref{sec:security}), and significantly better accuracy 
(Section~\ref{sec:eval}).

We empirically evaluate proximal consensus against a
convex hull method based on byzantine vector consensus~\cite{byz-vector-consensus} by
simulating their performance under varying noise distributions and
network characteristics.
Our results demonstrate that the output values produced by proximal consensus 
are more accurate than \emph{vector consensus}, a common convex hull protocol, 
across varying levels of noise.
Specifically, we
show that proximal consensus decreases the median percent error 
in comparison to
vector consensus
by an average of $56\%$ to $65\%$ under no attacks.
When the system is under an optimal Byzantine attack, defined in 
Definition \ref{def:effective-attacks},
we show that proximal consensus decreases the median percent error by
$31\%$ to $78\%$. 
In both scenarios we simulated a replica set with $f=1$ to $f=4$
Byzantine replicas when 
noise on the scale of $2\%$ to $12\%$ is introduced to the system.
Our results are statistically significant at a confidence level
of $99.9\%$.

%% file: PC.tex
\section{Proximal Byzantine Consensus}
\label{sec:pc}

We define the \emph{proximal Byzantine consensus problem} below:
\begin{definition}[Proximal Byzantine Consensus Problem]
  \label{def:bpcp}
  A node receiving outputs $Q$ from non-faulty replicas 
  modeled by random variable $X \sim D$ over a channel with
noise modeled by $Y \sim D_{\epsilon}$ decides on a value $v$ 
with confidence $c_{X\cdot Y}$ such that:
{\small
\begin{align}
v \in \bigl[ E( X \cdot (Y - \epsilon_L ) \mid Q),\; E(X \cdot (Y + \epsilon_H ) \mid Q) \bigr] \notag
\end{align}}
where $Y-\epsilon_L$ and $Y+\epsilon_H$ is the negative and positive range of the error, and $c_{X\cdot Y}$
is the confidence over that range.
\end{definition}

In this paper, we give a solution for the (common) case where $X$ and $Y$ are
independent and the noise distribution is unbiased, meaning the expected value
of $Y$, $E[Y]$, is 1.\footnote{When $E[Y] \neq 1$, it implies the noise causes a
skew in the observed outputs. In this case, our $E[Y] = 1$ assumption just
causes consensus values to be similarly skewed.} The observations in $Q$ are
nominally derived from the same system input streams---each replica (attempts
to) receive the same stream of outputs from a sensor or upstream replica set.
However, since some replicas may experience network faults, we model the
observed outputs of each replica scaled by an independent noise sample $Y$. 

Rather than choosing interval parameters $\epsilon_L$ and $\epsilon_H$ directly as in other
approximate byzantine consensus protocols, the system designer selects the required confidence $c_{X \cdot Y}$,
and $\epsilon_L$ and $\epsilon_H$ are derived from  $c_{X \cdot Y}$ and the variance of output observations.
Higher values of $c_{X \cdot Y}$ result in wider intervals, as do output observations with higher variance.
Informally, the noisier non-faulty values are, the more difficult it is to distinguish non-faulty outliers from 
Byzantine values,  so we want the smallest 
values of $\epsilon_L$ and $\epsilon_H$ with sufficient confidence for the variance we seek to tolerate.

Here we exploit the fact that each observation in $Q$ is the product of
a single sample from $X$ and multiple independent samples from $Y$ to directly
infer the error distribution parameters from the observed replica outputs, resulting
in interval bounds $\epsilon_L$ and $\epsilon_H$ that are proportional to the
variance of $Y$. However, a trivial variant of our solution also applies to systems where, rather than
consuming the same input stream, the replicas consume independent,
identically-distributed (IID) input streams. In this case, error distribution
parameters cannot be inferred directly, so the bounds would be proportional to the 
variance of $X \cdot Y$.  For simplicity, we only discuss the former configuration in the 
remainder of the paper.

Since
$X$ and $Y$ are also conditionally independent with respect to $Q$, 
the interval of $v$ simplifies to
{\small
\begin{align}
v \in \bigl[ E( X \mid Q) \cdot (1 - \epsilon_L),\; E(X \mid Q) \cdot (1 + \epsilon_H)  \bigr] \notag
\end{align}}
In our solution presented in Section~\ref{sec:pc-details}, we assume $X$ and $Y$ have stationary Gaussian 
distributions, but expect our approach is applicable to other distribution families.

Like other fault-tolerant protocols, proximal consensus relies on a
majority of non-faulty replicas overwhelming the behavior of faulty
replicas. What sets it apart from exact BFT protocols is that
proximal consensus ($\PC$) enables an \emph{approximate} (or ``proximal'')
consensus to be formed when, due to network faults or corrupted
or crashed replicas, non-faulty replicas produce differing outputs based on
incomplete or otherwise corrupted data. Furthermore, Proximal Byzantine Consensus is distinct from approximate consensus
problems such as \emph{asymptotic agreement}~\cite{LeBlancHKX11},
\emph{approximate scalar agreement}~\cite{DolevLPSSW86}, and \emph{approximate vector agreement}~\cite{byz-vector-consensus,multi-dim-vc} 
in that neither the problem nor the solution are defined in terms of the convex hull of non-faulty replica outputs.

The core idea behind proximal
consensus is that even when
the outputs of
non-faulty nodes differ, these 
differences can be modeled statistically to account for the noise 
introduced into the replicated computations.
Specifically, proximal consensus attempts to determine the
subset of replica outputs most likely to have
been produced by non-faulty replicas. For example, if we have $n$ replicas
and believe at most $f$ of them could be faulty or malicious, then we want
to determine the most likely subset of outputs of size $2f+1$. This ensures
that a majority of outputs in the subset came from non-faulty replicas.
Given that subset, we want to determine the most likely
\emph{ideal output}: what the output would be if the nodes received all
messages in time to process them.  To determine this ideal output, the
nodes receiving the replica outputs consider the conditional
probability of receiving each subset of $2f+1$ outputs from non-faulty nodes given the
current statistical model of the output domain scaled by a noise factor.

%% file: system_model.tex
\section{System design and threat model}
\label{sec:system-model}

\begin{figure}
    \centering
    \includegraphics[width=0.6\linewidth]{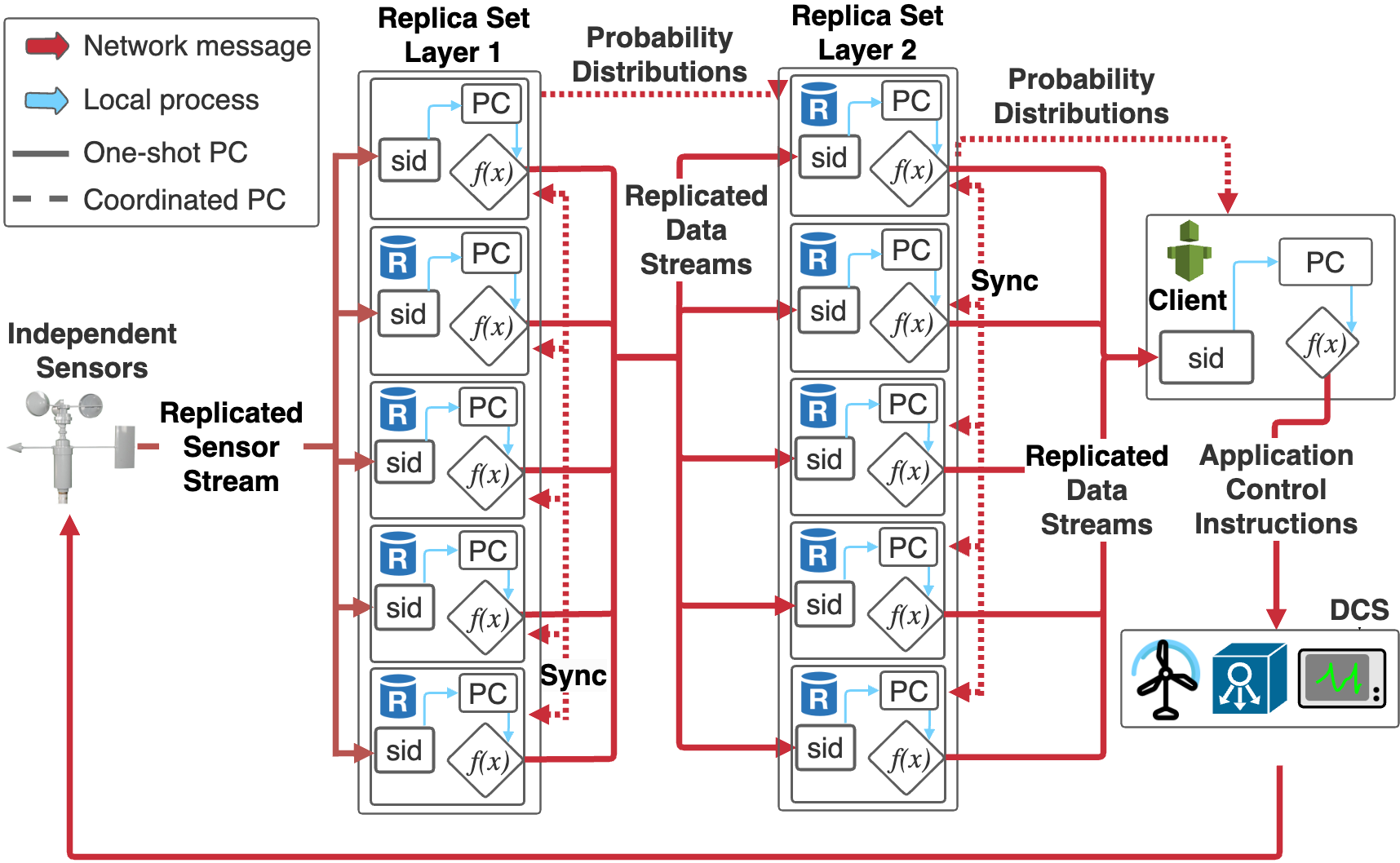}    
    \caption{System Architecture for a fault-tolerant data pipeline.}
    \label{fig:architecture}         
\end{figure}

\paragraph{\textbf{One-shot proximal consensus}}
When timely outputs are prioritized over consistent ones, the ``one-shot'' variant
of our proximal solution provides additional flexibility over existing
approximate BFT protocols. These BFT protocols incorporate a convergence parameter
$\epsilon$ that could in theory produce quicker convergence on a consensus
output, but may still require multiple rounds to terminate. Furthermore, it is
difficult to assess the application-specific impact of larger values of
$\epsilon$ since the area of convergence in the convex hull may be largely under
the control of the attacker.

In one-shot proximal consensus (Algorithm~\ref{alg:one-shot-prox} in Appendix), clients receive outputs directly from 
upstream replica sets 
and calculate the proximal consensus value locally 
using Definition \ref{PC-definition}.
Given the set of received values, the client
calculates its confidence in the candidate consensus value.
If the confidence
exceeds a minimum threshold, it accepts the value and may act on it. Otherwise, a
network fault may have interrupted message delivery,
and the client must wait for additional messages, up to a total of $n-f$.
If $n-f$ messages have been received without meeting the minimum confidence threshold, 
the client accepts the consensus value, but may choose not to act on it based on 
the low confidence.\footnote{Whether or not to act on low-confidence values is 
somewhat application specific. For example, it may be preferable to respond quickly 
if the currently-received messages result in high confidence. Otherwise 
a client could wait for the ``best possible'' confidence, and act on the value 
as long as it exceeds a lower confidence threshold.}

Since each client computes a value based on its own view of
the replica outputs, the output values and intervals may differ between clients.
However, since all intervals contain the ideal output with high probability, 
they (with high probability) overlap with the interval determined by a 
hypothetical client that successfully received all outputs. 
One-shot proximal consensus is particularly appropriate for
replicating approximate data-stream computations where exact agreement between 
clients is unnecessary and occasional dropped or low-confidence values are
tolerable.

\paragraph{\textbf{Coordinated proximal consensus}} Instead of having clients compute
proximal consensus values themselves, coordinated proximal consensus 
(Algorithm~\ref{alg:coordinated-prox} in Appendix) reaches a
consensus value among the replicas before providing a result to the client.
This workflow is closer to traditional consensus protocols and eliminates the 
case where clients may reach different decision values due to observing
different subsets of replica outputs. Using a traditional consensus protocol,
replicas first reach agreement on a set of (possibly different) outputs
received from their peers. They each then deterministically
compute the proximal
consensus value and confidence interval, and send the results to the client.
Since the replicas observe the same stream of observations, the updated distribution parameters of non-faulty 
replicas remain consistent.

In a \textit{\textbf{hybrid configuration}}, a coordinated instance of proximal consensus can be used 
to correct temporary divergence between one-shot instances by periodically sending updated
distribution parameters to its consumers (i.e., any clients computing one-shot consensus).

\paragraph{\textbf{System model}}  Figure~\ref{fig:architecture} illustrates a hybrid configuration
of proximal consensus for the feedback
dataflow of a distributed control system (DCS). Edges
represent network connections between distributed components.
Solid lines represent the one-shot dataflow path, which acts as a speculative
``fast path'' enabling consumers to receive a set of outputs with minimal 
dependency on inter-replica coordination.
The dashed lines represent the additional processes of the coordinated dataflow path.
These processes maintain 
a consistent set of inferred distribution parameters agreed upon by the replicas within the replica set.
By periodically updating the downstream one-shot replicas with these parameters, the degree
and duration of divergence between one-shot replicas is reduced.

System components that process, clean, and store data from upstream
producers are called \emph{consumers}. The end-user client in
Figure~\ref{fig:architecture} is a consumer, but so are the replicas
in the first and second layers.  
System components that produce and stream data to
downstream components are called \emph{producers}. For example, the sensors on
the left hand side of Figure~\ref{fig:architecture} are producers, but
the replicas in the first and second layers are also producers.

\paragraph{\textbf{Network model}} 
%
%

Proximal consensus is designed for robustness to both Byzantine and
network faults, so asynchronous or partially synchronous
settings are the most interesting since messages from sensors or non-faulty replicas
may not arrive by the time a
downstream consumer wishes to act. In a synchronous network,
missing messages are considered faults, which reduces the ambiguity between
faulty and non-faulty behavior since ``noise'' is attributable to faults.
Synchronous proximal consensus is more interesting for the variant (see
Section~\ref{sec:pc}) where replicas consume IID input streams rather than
subsets of the same stream, where noise may be due to the variance of the input
streams or faulty behavior. 
We focus primarily on the non-synchronous case (asynchronous or partially
synchronous) where, due to network delays, non-faulty node computations
experience some noise when they must produce outputs based on a subset of stream
inputs.

\paragraph{\textbf{Threat model}} 
Each replica set $R_{i}$ contains $n_i \geq 4f_i+1$ replicas with up to $f_i$
being Byzantine. Each round of proximal consensus produces a value $v$ and an interval guarantee
$IG=[v-\widehat{\epsilon_L}, v+\widehat{\epsilon_H}]$, where
$\widehat{\epsilon_L}$ and $\widehat{\epsilon_H}$ are derived from the current parameters
inferred for $Y$. The $IG$ indicates an interval around $v$ that contains the true output
with high probability, and represents the uncertainty of $v$ relative to the
true output. 

Replica sets and one-shot clients are configured with a confidence level $c_i$
and an acceptable interval width, $AIW_i$.  The $c_i$ parameter defines what
confidence is required for the interval guarantees of each proximal
consensus decision.  The $AIW_i$ parameter is used to reduce consensus latency
by specifying the (desired) minimum acceptable width of the interval defined by
$\epsilon_H$ and $\epsilon_L$ for confidence level $c_i$. 

Attacker influence is bounded by ensuring that any attack included in a selected
quorum (and therefore able to influence the output) is at least as likely to have
come from a non-faulty replica as the values that were \emph{not} selected.  
The key to providing this guarantee is that a sufficient number of messages must
be received so that a $2f_i+1$ quorum can be selected out of
minimum $3f_i+1$ messages. Quorums need at least $2f_i+1$ outputs
so that even in the worst case, when the only non-faulty quorum is the least likely one
considered possible (e.g., containing outputs within the 99.7\% confidence interval),
the attacker's influence is still bounded by the interval guarantee. Intuitively, 
the \emph{worst-case honest quorum} is split between extremal values, 
giving the attacker maximal influence. However,
since each member of the selected quorum must be at least as likely as the 
(at least) $f_i$ unselected messages, any attacks that are selected for the quorum
must be as likely as an honest output that was not selected. We evaluate these 
guarantees formally in Section~\ref{sec:security}.

A quorum can also be selected when the resulting interval guarantee exceeds the
$AIW_i$ parameter.  Essentially $AIW_i$ acts as a proxy for unseen honest
outputs, declaring an application-specific degree of influence that is considered
tolerable.  If a $2f_i+1$ quorum exists among the received outputs that
results in an acceptable interval guarantee, then it is unnecessary to wait for
$3f_i+1$ messages.

The need for $3f_i+1$ messages in the general case increases the minimum number
of replicas to $4f_i+1$ for proximal consensus compared to convex hull
approaches, most of which require only $3f_i+1$. To ensure liveness while
preserving the soundness of the interval guarantee, it is necessary that a
replica can expect to eventually receive $3f_i+1$ messages.  This is only
possible if there are at least $3f_i+1$ non-faulty replicas since up to $f_i$
Byzantine replicas could withhold outputs indefinitely. This also implies that 
in synchronous networks,
as long as the upper-bound
on message delivery is shorter than the desired reaction time,
only $3f_i+1$ replicas are required since non-faulty
replica outputs always arrive on time.

%% file: implementation.tex
\section{Consensus as probability maximization}
\label{sec:pc-details}

\begin{wrapfigure}{r}{0.5\textwidth}
    \centering
    \includegraphics[width=.41\textwidth]{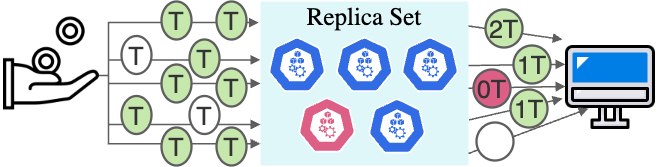}
    \caption{Counting the number of tails flipped.}
    \label{fig:coinflip}
\end{wrapfigure}

We present the mathematical foundations of proximal consensus using a
simple example. Consider a stream of coin flips
sent to a replica set
that produces the observed count of tails results for the past $t$ seconds.
In Figure~\ref{fig:coinflip}, the data source flips a
coin twice and sends each result (both tails) to the replicas.
Two non-faulty replicas and one Byzantine replica receive both observations. 
Although the Byzantine replica receives both messages, it 
outputs \texttt{0T}, trying suppress the total tails count.
The two non-faulty replicas send \texttt{2T}, but only one output arrives at 
the client due to a network partition.
Two other non-faulty replicas receive only one of the observations, 
thus outputting \texttt{1T} to the client.
From the clients view, any one (but only one) of the received values may be faulty,
or they could all be non-faulty and the missing value may be from a faulty replica.

Exact BFT protocols require a majority (typically two-thirds) of replicas to
agree in order to decide on a value. In the Figure~\ref{fig:coinflip} scenario,
no such majority exists.  The client instead uses proximal consensus to
determine the most likely number of tails given the outputs it received and its
prior knowledge, if any, about the coin flipping process.  It models each coin 
flip as a sample from a uniform distribution of heads/tails, scaled by a
randomized noise factor that accounts for error introduced by network latency or
other transient sources of noise experienced by the client receiving
inputs.

First consider the unconditional probability of receiving \texttt{2T}. A
non-faulty replica reports $2T$ if it observes two \texttt{T}
messages from its source. Therefore, the probability of observing \texttt{2T} 
is the joint probability of two tail flips and two delivered messages.
The joint probability of two tail flips is $P(TT)=\frac{1}{2}\cdot \frac{1}{2}$. 
Now suppose $R_1$ and $R_2$ represent
random variables for arrival times of the first and second messages at
the replica. The joint probability of two delivered
messages is $P(R_1\leq t) \cdot P(R_2 \leq t)$
where $t$ is the deadline. This joint probability models the ``noise''
introduced by observing coin flips over the network.
Assuming $R_1$ and $R_2$ are independent of
the contents of the messages, the joint probability of two tail flips 
and two delivered messages is
$P(2T)=$ $\frac{1}{4}\cdot P(R_1\leq t) \cdot P(R_2 \leq t)$.

Now consider receiving \texttt{1T} from an honest node. Here,
more than one scenario could have produced this output.  There could
have been only one tail: either \texttt{TH} or \texttt{HT} with the
\texttt{T} message delivered, or there could be two tails
(\texttt{TT}) but one of the messages failed to arrive in time.
The following sum of joint probabilities represent these scenarios.
{\small
\begin{align*}
    P(1T) &= P(TH) \cdot P(R_1 \leq t) + P(HT) \cdot P(R_2 \leq t) + P(TT) 
    \cdot \big(P(R_1 > t) \cdot P(R_2 \leq t) + P(R_1 \leq t) \cdot P(R_2 >t)\big)
\end{align*}
}

\normalsize
Receiving \texttt{0T} from an honest node could result
from two heads or all tails messages being delayed.
{\small
\begin{align*}
    P(0T) &= P(HH) + P(TT) \cdot P(R_1 > t) \cdot P(R_2 > t)    
    + P(TH) \cdot P(R_1 > t) + P(HT) \cdot P(R_2 > t)
\end{align*}
}

\normalsize
\noindent Using these probabilities and the messages received from replicas, proximal consensus aims 
to determine the most likely \emph{ideal output}: 
what a non-faulty node would output if there were no network or
replica faults. 
Formally, let $X$ be a random variable representing
the sample space of the ideal output
and $x$ be a candidate output in the (ideal) output sample space
($x \sim X$),
and $q_i \in Q$ be samples of the replica output sample space ($q_i \sim XY$) where $Y$ 
is the noise distribution.
Let $q$ be a subset of size $2f+1=3$ of the
observed events $Q = \{q_1=2T, q_2=1T, q_3=0T, q_4=1T\}$. The client's task is
to find a pair $(x,q)$ 
that maximizes
the conditional probability
$P(X=x ~|~ q)$:

\begin{definition}[Proximal consensus of $Q$]
\label{PC-definition}
{\small
  \[
  \PC(Q,n,f) \triangleq \argmax_{x \in X \;;\; q \in [Q]^{2f+1}} ~ P(X = x ~|~ q )
  \] 
}
\end{definition}
In other words, the client finds the most likely output 
$x$ from the output sample space $X$
over all valid quorums $q$ of size $2f+1=3$.
This pair $(x, q)$ maximizes the probability $P(X = x ~|~ q)$.\footnote{In 
Bayesian statistics, this process is often referred to as computing the 
\emph{maximum a posteriori} or \emph{highest posterior density interval (HPDI)}.}
It will sometimes be convenient to talk about the
most likely output $x$ resulting from a fixed quorum $Q$,
which we will refer to as $x = \PC(Q, n, 0)$.

The observations during a round of proximal consensus are used to improve the
inferred distribution parameters for subsequent rounds. Currently we consider
analytical distribution families with known conjugate prior
functions~\cite{bayes-textbook}. These functions update the prior distribution
parameters to form the \emph{posterior distribution}, which takes into account the messages observed in the
selected PC quorum.  By only considering the selected PC messages, we reduce
influence from faulty values at the time of inference.

The inference process infers \emph{distributions} on the
unknown distribution parameters rather than a single ``best'' choice of
parameters---the values observed could have been produced by one
of many potential source distributions.  To choose the most likely output value,
we must consider all of the potential distributions, weighted by the probability
they could have produced the selected observations.  Fortunately, for
distribution families with conjugate prior functions, this task is
straightforward using the \emph{posterior predictive distribution}~\cite{bayes-textbook}, which gives
the likelihood of an output given the posterior distribution of the parameters.

\paragraph{\textbf{Computing a proximal consensus}} Like vector consensus,
the consensus value $x=\PC(Q,n,f)$ may not necessarily be any of the outputs in $Q$. Since the outputs
in $Q$ are neither independent of the distribution they were produced from nor each other, we can
expect non-faulty outputs to be similar in relatively reliable networks. 
Consequently, we can compute the conditional probability $P(X=x~|~q)$ based on a similarity function proposed by 
Blok et al.,~\cite{similarity-prob} that relates the replica outputs and the candidate event:
\begin{definition}[Conditional Probability]
\label{def-conditional-probability} \label{eq-cond-prob}
\small{
\begin{equation}
        P(X=x~|~q) = P(x)^\alpha \quad
        \text{where } \alpha = \frac{1-\similar([x,q])}{1+\similar([x,q])}^{1-P(q)}  \notag 
\end{equation}}
\end{definition}
For example, as the similarity of $x$ with the $2f+1$ observations in $q$ 
($\similar([x,q])$) approaches 1 (more similar), 
$P(x)^{\alpha}$ approaches 1. 
Thus, $\alpha = 0$ means $x$ matches every output in $q$ or $P(q)=1$.

Using Definition~\ref{def-conditional-probability}, proximal consensus computes the global maxima point of each quorum $q\in Q$
and returns the $(x,q)$ pair that is most conditionally likely.
Finally, the currently inferred mean of the replica output distribution ($\mu_{XY} = E[XY]$) and 
variance of the error distribution ($\sigma_\epsilon = \sqrt{\text{Var}(Y)}$) 
are used to conservatively estimate a 99.7\% interval guarantee (IG) for the 
chosen value: 
$IG = \bigl[\mu_{XY} \cdot (1 - 3\cdot \sigma_\epsilon) , \mu_{XY} \cdot (1 + 3\cdot \sigma_\epsilon)\bigr]$.
Algorithm~\ref{alg:prox-consensus} and the mathematics of a proximal consensus (Equations~\ref{eq-conjugate-prior}-\ref{eq-2d-dist}) 
is provided in the Appendix.

\subsection{Run-time complexity}
\label{sec:complexity}
In one dimensional convex hull methods, computing valid consensus points given a set of 
replica outputs (whether Tverberg points or
centerpoints~\cite{vc-centerpoint}) is equivalent to finding the median, and so 
has time complexity of $O(f\log f)$.\footnote{Although in exact BFT protocols there is usually no reason to choose
an $n$ higher than the minimum (e.g., $n=3f+1$, implying $O(n) \sim O(f)$), $n > 3f+1$ can improve accuracy in our setting for both vector and proximal consensus. For this 
reason, we state asymptotic complexities in terms of $f$ when they involve the size of the quorums needed to tolerate $f$ faults, and use $n$ only 
when it involves the total number of replicas.}  In higher dimensions $d$, the complexity
rapidly increases.  Currently-known Tverberg point algorithms have complexity
$O((2f+1)^{d(d+1)+1})$~\cite{tverberg-median}.  Based on prior work establishing that
checking the validity of centerpoints is co-NP-complete~\cite{centerpoint-coNP},
Abbas et al.~\cite{vc-centerpoint} conjecture checking the validity of consensus 
points in convex hull methods is also co-NP-complete.

Approximate algorithms for computing
Tverberg points exist~(see \cite{tverberg-median} for an overview), 
but require more non-faulty nodes per tolerated faulty node: $f < \Omega(\frac{n}{2^d})$ for Tverberg
points, and $f < \Omega(\frac{n}{d^2})$ for centerpoints~\cite{vc-centerpoint}.
Approximation parameters for these algorithms that
guarantee fault tolerance~\cite{vc-centerpoint} imply a runtime complexity of 
$$O((2f+1)^{c \log d} \cdot (2d)^d)$$ 
More recent approximation algorithms~\cite{tverberg-median}
improve the runtime complexity in general, but it is not currently clear to what
degree they improve the efficiency of valid consensus point calculations.

Convergence in vector consensus depends to some extent on computing valid 
consensus points for each potential $2f+1$ quorum of the received values, 
implying that the above calculations are performed up to $n \choose 2f+1$ times
per round.  Most vector consensus algorithms converge quickly, but it may take
multiple rounds to decide a value. Assuming convergence occurs in a constant
number of rounds, the computational complexity of agreement with 
vector consensus is $O({n \choose 2f+1} \cdot f\log f)$ for one dimension, and 
for arbitrary dimensions (using approximate Tverberg point algorithms):
 $$O({n \choose 2f+1} \cdot (2f+1)^{c \log d} \cdot (2d)^d)$$

Proximal consensus also considers each of the up to $n \choose 2f+1$ quorums, 
but the complexity of the calculations performed for each quorum scales
significantly better.  To compute the $\argmax$ with respect to a quorum $q$ and
consensus value $x$, our solution determines the $x$ with maximum likelihood for
a given $q$ using a binary search over a search domain $S$ quantized by a step 
size $p$. At each candidate $x$, evaluating the conditional probability involves
computing the similarity score~\eqref{eq:similarity-function} of $x$ with each of the $2f+1$ quorum outputs, 
which is based on a normalized $d$-dimensional Euclidean distance (see \eqref{eq-2d-dist}). 
The normalization step is $O({2f+1 \choose 2}) \sim O(f^2)$~\cite{erdos-graphs} and the distance calculation is $O(d)$.
Thus each iteration of the search algorithm is $O(f^{3} d)$.
The search requires up to $O(\log k)$ iterations where $k =
\frac{|S|}{p}$ i.e., the number of intervals $S$ in the search space implied by
step size $p$.\footnote{To further improve efficiency
our implementation limits the search to the the 
99.7\% credible interval of the distribution, which significantly reduces the 
size of $k$ relative to step size.}

This paper presents a proximal consensus approach for one dimensional values.
While our approach needs further study to evaluate its effectiveness in $d$
dimensions, the complexity of the $\argmax$ calculation in $d$ dimensions is
easy to analyze.  For general dimensions $d$, finding the maximally likely value
for $x$ in a uni-modal distribution for a given $q$ can be accomplished with 
gradient descent (GD) algorithms, which have complexity $O(dk^2)$. Furthermore,
stochastic gradient descent (SGD) algorithms significantly improve performance
by approximating the gradient calculation. SGD algorithms are also less likely
to converge on local maximums, and so may be appropriate for multi-modal
distributions. 

In summary, the asymptotic complexity of our proximal consensus algorithm is
$O(f \log k)$ and for $d$ dimensions is expected to be (using exact gradient 
descent search):
$$O({n \choose 2f+1}\cdot f^3 d^2 k^2)$$
This indicates that although the runtime complexity of our algorithm has 
comparable or even slightly worse asymptotic complexity ($k$ will typically 
be larger than $f$) in one dimension, it is significantly more efficient 
in higher dimensions than convex hull-based protocols.

%% file: impact.tex
\subsection{Security guarantees against adversaries}
\label{sec:security}
\newcommand{\NIG}{\ensuremath{N\!-\!\Gamma^{-1}}}
\newcommand{\XY}{\ensuremath{X \cdot Y}}

\RS{Reminder to update the dist and sim functions in this section if Owen likes the changes
in the above section.}

We now present a detailed analysis of the maximum impact of Byzantine
replicas on a replicated data stream. 
Byzantine adversaries are omniscient and adaptive in the sense that they have
complete knowledge of the outputs of all correct replicas prior to
choosing their malicious outputs to maximize the effectiveness of the
attacks. 
As discussed in Section~\ref{sec:system-model}, for an attack to be effective, 
it must be likely enough to be a non-faulty output that it is included in the selected quorum.
We formally define these attacks below:
\begin{definition}[Effective attacks]
  \label{def:effective-attacks}
  Let $Q= \{r_1, ..., r_{(2f+1)}\}$ be a quorum containing $2f+1$ outputs from correct replicas,
  $Q_L \subset Q$ be the smallest
  $f+1$
  honest outputs, and $Q_H \subset Q$ be the largest 
  $f+1$
  honest outputs.
  Furthermore, define 
  $A_L$ (resp. $A_H$) to be a set of $f$ Byzantine outputs aimed to suppress (inflate) the $\PC$ output.
  Let $\phi_L = Q_L \cup A_L$ and $\phi_H = Q_H \cup A_H$.
  A \textbf{\emph{suppressing attack}} is a set $A_L$ such that
  {\small
  \begin{align}  
      & \PC(Q, 2f+1, 0) > \PC(\phi_L, 2f+1, 0) ~~
    \text{and} ~~  P(X=\PC(Q, 2f+1, 0)~|~Q)   
        \leq  P(X=\PC(\phi_L, 2f+1, 0)~|~\phi_L) 
  \end{align}}
  \normalsize
  An \textbf{\emph{inflating attack}} is a set $A_H$ such that
  {\small
  \begin{align}  
      & \PC(Q, 2f+1, 0) < \PC(\phi_H, 2f+1, 0) ~~
   \text{and} ~~ P(X=\PC(Q, 2f+1, 0)~|~Q) 
       \leq P(X=\PC(\phi_H, 2f+1, 0)~|~\phi_H)       
  \end{align}}
  \end{definition}
  Satisfying assignments for $A_L$ and $A_H$ always exist by
  choosing $A_L$ such that $Q_L\cup A_L = Q$ or $A_H$ with
  $Q_H \cup A_H = Q$. We call the $A_L$ containing the lowest
  satisfying outputs a \emph{maximally suppressing attack} and the
  $A_H$ containing the highest satisfying outputs a \emph{maximally
  inflating attack}.



Let $X \sim N_{\chi}(\mu, \sigma)$ and $Y \sim
N_{\epsilon}(1,\sigma_{\epsilon})$ be independent random variables for,
respectively, the output distribution of data stream $\chi$ and the error
distribution for the network channels $\chi$ is broadcast on.
The mean $\mu$ and variance
$\sigma^2$ of $X$ is unknown, as well as the variance
$\sigma_{\epsilon}^2$ of the error distribution.  Since the error distribution is
unbiased, we set the mean $\mu_{\epsilon}$ of the error distribution to
$\mu_{\epsilon}=1$. 
We take a Bayesian approach to modeling the unknown
parameters of these distributions: $\mu, \sigma, \sigma_{\epsilon}$ are
random variables.  
Outputs of size $n$ replica sets are modeled using a sample $x \leftarrow X$,
and between $3f+1$ and $n$ samples $y_i \leftarrow Y$ to give outputs $xy_1,
xy_2, ..., xy_n$.  Only these samples are directly observable by 
consumers, and are approximately (cf.~\cite{compute-product-random-variable}) distributed 
as $XY \sim
N_{\chi\epsilon}\left(\mu, \sigma_{XY}=\sqrt{\sigma^2+ (\mu^2+\sigma^2)\cdot \sigma_{\epsilon}^2}\right)$.

Therefore, our goal is to use observations from the replica outputs to infer the
parameters $(\mu, \sigma_{XY})$ of $N_{\chi\epsilon}$,
which has a normal-inverse-gamma (\NIG)
prior distribution, and the error distribution variance $\sigma_{\epsilon}$, which has 
 an inverse-gamma distribution.
Our guarantee takes the form of a bound on the influence
Byzantine replicas may have on the output of a proximal consensus round
given an error distribution. Since the error distribution varies
(inversely) with the similarity between non-faulty outputs,
our guarantee is parameterized on a prediction interval (PI) for $N_{\chi\epsilon}$.

The selected PI bounds the sample space of observed outputs
under consideration: outputs from non-faulty replicas falling 
outside the PI are considered too
unlikely to affect the ability of an attacker to do harm.  For
example, we might wish to consider all non-faulty outputs occurring within
the $99.7\%$ PI. Consequently, if the \emph{most probable honest
quorum} of a round contains outputs outside this interval, the
stated bound could be violated.  However, for even one of
this quorum's $2f+1$ outputs to fall outside the $99.7\%$ PI, all
other honest outputs not in the quorum must also fall outside the PI.
If $n=4f+1$, then only $2f+1$ received outputs are guaranteed to be
non-faulty since $f$ may be missing and $f$ of the $3f+1$ received messages
may be faulty. Therefore, the chance of one of these honest outputs falling
outside the PI is $0.3\%^{(2f+1)}$ since each output represents an independent sample
from the error distribution $N_{\epsilon}$.  For $n > 4f+1$, the probability of
a non-faulty output in the selected quorum falling
outside the PI decreases as $n$ increases.

In terms of Definition~\ref{def:bpcp}, let $v \in
[v_{\text{low}},v_{\text{high}}]$ be the interval specifying the
bounds on the value $v$ decided by a client given a set of received
replica outputs, and let $c_{\chi\epsilon}$ be a percentage
representing a credible interval $\mathcal{I}_{\chi\epsilon}$
on the observed output distribution $N_{\chi\epsilon}$.  The
attacker is only able to cause the consumer to decide a value $v'
\not\in [v_{\text{low}},v_{\text{high}}]$ if all but $2f$ outputs from
honest nodes fall outside $\mathcal{I}_{\chi\epsilon}$.
Therefore, the confidence level $c_{\epsilon}$ that $v$ is bounded by
$[v_{\text{low}},v_{\text{high}}]$ is equal to the probability of
not receiving at least $(n-f)-2f$ honest outputs
outside $\mathcal{I}_{\chi\epsilon}$, thus
\begin{equation}
\label{eq-conf}
   c_{\epsilon} = 1 - (1-c_{\chi\epsilon})^{(n-3f)}
\end{equation}

\normalsize
Notice that the attacker cannot affect this probability since it
depends only on the 
similarity of
honest outputs, which affects the 
$\mathcal{I}_{\chi\epsilon}$.  Furthermore, the above formula
represents the probability of the bounds being violated \emph{at all}.
The larger the security bound violation, the less
likely the consumer is to receive honest outputs that
are sufficiently
dissimilar
to enable the attack.

To determine the bounds $v_{\text{low}}$ and $v_{\text{high}}$, we
consider the worst-case honest quorum for the output credible
interval $\mathcal{I}^{\text{obs}}_{\chi}$. These outputs give the
attacker maximum power to influence the proximal consensus result. We
consider the outputs in this quorum possible, hence in
$\mathcal{I}_{\chi\epsilon}$, but potentially very unlikely. In
fact, we want to consider the \emph{lowest probability} honest quorum,
since minimizing $P(X = \PC(Q, 2f + 1, 0) ~|~ Q)$ in
Definition~\ref{def:effective-attacks} increases the attacker's
potential choices for $A_L$ and $A_H$.  Therefore, the maximum
influence an attacker has is 
a maximally inflating (or suppressing) attack when the outputs of the
honest nodes form a worst-case honest quorum.

We determine the worst-case honest quorum as follows.  Observe that
$P(X = \PC(Q, 2f + 1, 0) ~|~ Q)$ in
Definition~\ref{def-conditional-probability} is minimized when the
similarity $\similar({[X,Q]^{\textsf{norm}}})$, defined in \eqref{eq:similarity-function}, is minimized, 
which in turn is minimized when the normalized distance $\dist(D=[X,Q]^{\textsf{norm}})$, 
defined in \eqref{eq-2d-dist}, is
maximized. Therefore, the worst-case honest quorum is an honest quorum
of outputs in $\mathcal{I}_{\chi\epsilon}$ with a maximal
normalized distance. 
The maximum distance between two honest outputs depends on the size of 
the credible interval $\mathcal{I}_{\chi\epsilon}$. For $c_{\chi\epsilon} = 99.7\%$, 
the interval is
$\mathcal{I}_{\chi} = \bigl[\mu \cdot (\mu_\epsilon - 3\cdot \sigma_\epsilon)
                                      , \mu \cdot (\mu_\epsilon + 3\cdot \sigma_\epsilon)\bigr]$.
                                      
A maximal normalized distance, then, is a quorum with outputs evenly
split between the upper and lower boundaries, $Q_H$ and $Q_L$ where
$Q=Q_H \cup Q_L$.  Since the quorum size is always odd ($2f+1$), there
are two choices depending on whether $Q_H$ or $Q_L$ is
larger. Which choice represents the worst case depends on the attack.
An attack ``replaces'' $f$ honest outputs in $Q$ with $f$ malicious
ones without increasing the normalized distance of the
quorum.  A suppressing attack has a greater effect if it can replace
all elements of $Q_H$, so the worst-case honest quorum $Q_S=Q^S_L \cup Q^S_H$ for a
suppressing attack is when $|Q^S_L|=f+1$ and $|Q^S_H|=f$. Dually, the worst-case
honest quorum $Q_I=Q^I_L \cup Q^I_H$ for
an inflating attack is when $|Q^I_L|=f$ and $|Q^I_H|=f+1$.

The maximally suppressing attack $A_L$ is then a set of $f$ outputs
$a_L$ where $a_L$ is the smallest value such that $\dist([A_L \cup
Q^S_L]^{\textsf{norm}}) = \dist(Q_S^{\textsf{norm}})$.
The maximally inflating attack $A_H$ is a set of $f$ outputs
$a_H$ where $a_H$ is the largest value such that $\dist([A_H \cup
Q^I_H]^{\textsf{norm}}) = \dist(Q_I^{\textsf{norm}})$.

We define $\Omega$ as the normalized Euclidean distance of the
worst-case honest quorum Q.
For $c_{\chi\epsilon} = 99.7\%$, 
\small{
\begin{align}
\label{eq:omega}
  \Omega           &= \begin{cases}
                    \frac{6\cdot\mu\cdot \sigma_\epsilon \cdot \sqrt{f\cdot(f+1)}}{\sigma} & \mu ≠ 0 \\
              \frac{6\cdot\sqrt{\sigma^2_{X\cdot Y}}\sqrt{f\cdot(f+1)}}{\sigma}              &  \text{ otherwise }
            \end{cases} \\
 \sigma^2_{X\cdot Y} &= (\sigma^2 + \mu^2)\cdot (\sigma^2_\epsilon + \mu^2_\epsilon)-\mu^2\cdot \mu^2_\epsilon
\end{align}}
\normalsize
Since $N_{\chi\epsilon}$ is symmetric, we have
{\small
\begin{align}
  a_L &= \mu \cdot (\mu_\epsilon - 3\cdot \sigma_\epsilon) - \dfrac{\Omega}{2} & 
  a_H &= \mu \cdot (\mu_\epsilon + 3\cdot \sigma_\epsilon) + \dfrac{\Omega}{2} \label{eq:a_H}
\end{align}
}

Therefore the worst-case suppressing impact $\Delta_S$ (for
$\mathcal{I}_{\chi\epsilon}$) is
{\small
$$\Delta_S=|\PC(Q^S, 2f+1,0) - \PC(A_L \cup Q^S_L,2f+1,0)|$$
}
\normalsize
and the worst-case inflating impact $\Delta_I$ is
{\small
$$\Delta_I=|\PC(Q^I, 2f+1,0) - \PC(A_H \cup Q^I_H,2f+1,0)|$$ 
}
\normalsize
These values are useful since for any other honest quorum $Q'$ and
maximally suppressing (or inflating) attack $A_L'$ ($A_H'$), 
the impact of these attacks on $Q'$ will be less than $\Delta_S$
and $\Delta_I$.

Since we know that $a_L \leq \PC(A_L \cup Q^S_L,2f+1,0)$ and $a_H \geq
\PC(A_H \cup Q^I_H,2f+1,0)$, we only need to estimate the proximal
consensus results for $Q^S$ and $Q^I$.  We observe
that by the definition of $Q^S$ and $Q^I$,
selecting the mean of each quorum as the result minimizes the distances
$\dist(D=[x,Q^S]^{\textsf{norm}})$ and $\dist(D=[x,Q^I]^{\textsf{norm}})$ between the result $x$ and each honest output in the quorum.

Although choosing the mean of the worst-case quorums minimizes $\dist(D=[x,Q^S]^{\textsf{norm}})$,
and therefore $\alpha$ (Definition~\ref{def-conditional-probability}), 
it may not maximize $P(x)^{\alpha}$. However, since $N_{\chi\epsilon}$
is uni-modal, the optimal $x$ will be somewhere in the interval between the mean of $Q^S$ (or $Q^I$)
and the mean of $N_{\chi\epsilon}$, $\mu \cdot \mu_\epsilon$. This allows us to create the following
bounds on the Byzantine impact:
{\small
\begin{align}
\Delta_S &\leq |\mu \cdot \mu_\epsilon - a_L| &
\Delta_I &\leq |\mu \cdot \mu_\epsilon - a_H|
\end{align}
}

\normalsize
\noindent From this we derive bounds for $\epsilon_L$ and $\epsilon_H$ (Definition~\ref{def:bpcp})
solely in terms of distribution parameters and system parameter $f$:
{\small
\begin{align}
  \epsilon_L &\leq \bigl|\mu_\epsilon - \frac{a_L}{\mu}\bigr| &
  \epsilon_H &\leq \bigl|\mu_\epsilon - \frac{a_H}{\mu}\bigr| 
\end{align}
}

\normalsize
\noindent Notice from Equations~\eqref{eq:omega}-\eqref{eq:a_H}, these bounds are
expressed solely in terms of distribution parameters and system
parameter $f$.

For a consumer of stream $\chi$ receiving
non-faulty outputs $Q$ and faulty outputs $Q_f$, where
$|Q| \geq 2f+1$, $|Q_f| \leq f$, and $n \geq 3f+1$, $\PC(Q \cup
Q_f,n,f) = v$ with confidence $c_\epsilon$~\eqref{eq-conf} such that
{\small
$$v \in \bigl[ E( X \mid Q) \cdot (1 - \epsilon_L),\; E(X \mid Q) \cdot (1 + \epsilon_H)  \bigr]$$
}

%% file: evaluation.tex
\section{Evaluation}
\label{sec:eval}

\begin{figure}
\centering
\begin{subfigure}{.32\textwidth}
  \centering  
  \includegraphics[width=\linewidth]{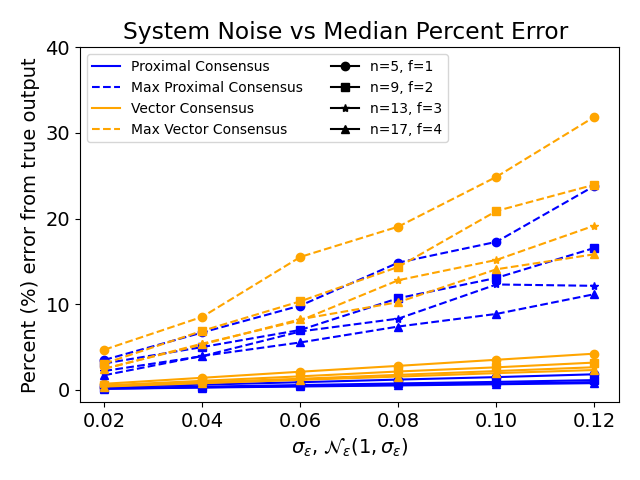}
  \caption{Error rates (no attacks).}
  \label{fig:no-attack}
\end{subfigure}
\begin{subfigure}{.32\textwidth}
  \centering  
  \includegraphics[width=\linewidth]{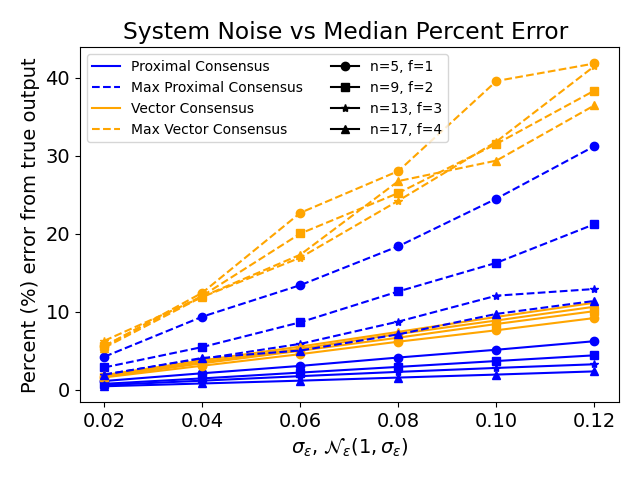}
  \caption{Error rates (optimal attacks).}
  \label{fig:byz-attack}
\end{subfigure}
\begin{subfigure}{.32\textwidth}
  \centering  
  \includegraphics[width=\linewidth]{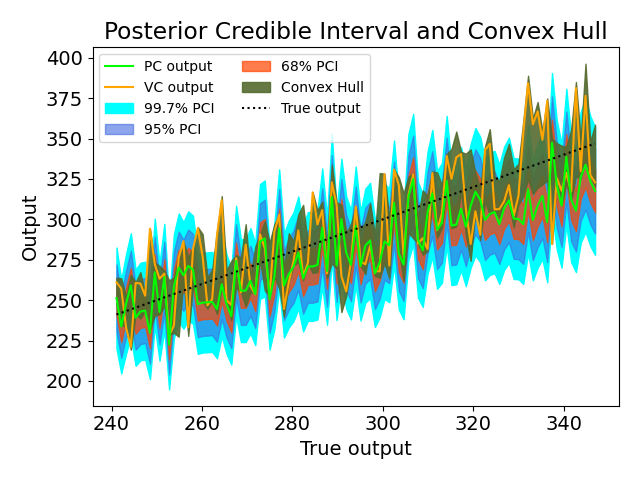}  
  \caption{PI (optimal attacks).}
  \label{fig:pc-predictive-interval}
\end{subfigure}
\caption{Evaluation results comparing the accuracy of PC and VC, as well as the PI and convex hull, respectively.}
\end{figure}

In this section, we evaluate the one-shot proximal consensus (PC) algorithm through
repeated simulations,\footnote{We focus on the one-shot PC variant; coordinated PC produces equal or better accuracy.} comparing it with the approximate
vector consensus (VC) algorithm from \cite{byz-vector-consensus} on 
one-dimensional vectors.\footnote{One-dimensional VC likely performs similarly to 
an approximate scalar consensus (e.g., \cite{approx-consensus,DolevLPSSW86}), but we chose VC since it generalizes
to higher dimensions.} 
We present two types of experiments, the first of which evaluates the median and maximum error under Byzantine
and no attack scenarios. The second experiment evaluates the coverage of the 
prediction interval (PI) of PC and the convex hull coverage of VC.
In all simulations,
we did not make any timing assumptions and assumed
the inter-replica network connection between VC replicas was synchronous. 
The first set of outputs provided by PC replicas was accepted regardless of how wide the resulting PI was. 
Results are statistically significant at a 99.9\% confidence level, with a level of precision (error margin) 
$e$ of $\leq 1\%$ from the presented outcomes.
The number of simulations $n_0$ to produce these results are calculated by the following equation: $n_0 = \frac{Z^2 \cdot \sigma^2}{e^2}$~\cite{sample-size},
where $Z=3.09$ is the $Z$-score and $\sigma^2$ is the variance of an attribute(s) 
in the population. Consequently, the number of experimental runs for each data point 
in our results varies between 209 for $\sigma_\epsilon=0.02$ to 4503 for $\sigma_\epsilon=0.12$.

Each simulation is comprised of one layer of $n=4f+1$ replicas where up to $f$ replicas
are Byzantine and a single client.\footnote{Although VC is presented as a $3f+1$ solution, for comparison purposes, the VC replica sets
also contained $4f+1$ replicas. Additional non-faulty replicas only improve the accuracy of VC compared to a $3f+1$ replica set.} 
In each interval of the no-attack experiment,
the client receives a set of $n$ outputs from replicas generated by multiplying a true output $x$ with $n-f$ randomly generated error samples $y$, such that replica outputs are $x\cdot y_1,\ldots x\cdot y_n$.
In the optimal attack experiment, the client receives $n-f$ outputs and an additional $f$ Byzantine outputs. 
We assume the strongest Byzantine attacker on both the PC and VC simulations that utilizes the noisy outputs of non-faulty replicas to maximally suppress or inflate the consensus output from the true output, whichever
results in a higher percent error from the true output.

In the first simulation, since the client does not have prior parameters of the output distribution a priori, the client first generates prior parameters by iteratively updating a prior distribution using the 
Gaussian conjugate prior function with unknown parameters~\cite{nig-conjugate} over five independent PC
rounds\footnote{These rounds were initialized with (uninformative) prior distribution parameters $\mu=294$, $\alpha=1$, $\beta=1$, $\nu=1$.} comprised exclusively of $n-f$ non-faulty replicas,
which the error of the following (sixth) consensus round being used as an error sample for iteration $i$ of $n_0$ iterations, calculated
using the expected standard deviation of $Y$ ($\sigma_\epsilon$).\footnote{We
conducted five iterative training rounds because our empirical evaluations under the specified configurations
resulted in distribution parameters from $n-f$ non-faulty outputs to closely match those from five iterative
training rounds with an additional $f$ Byzantine outputs.}

Figure \ref{fig:no-attack} and \ref{fig:byz-attack} shows the median and maximum PC and VC percent error from the true output
when under no Byzantine attack and Byzantine attack, respectively,
as the standard deviation of the error distribution increases. 
The plotted maximum error of PC also considered the error during the five rounds in which the prior distribution was generated. 
Overall, in the no-attack scenario, our results show that the one-shot PC variant reduces the median percent error of VC 
by $56\%-65\%$ and lowers the maximum 
percent error of VC 
by $23\%-30\%$ over $f=1\ldots 4$. 
In the Byzantine attack scenario, our results show that the one-shot PC variant reduces the median percent error of VC 
by $31\%-78\%$ and lowers the maximum 
percent error of VC 
by $31\%-68\%$ over $f=1\ldots 4$.

Figure \ref{fig:pc-predictive-interval} compares the output of PC and VC to the
true output after PC has converged on parameters over 500 rounds with 
non-faulty outputs sampled from $\XY$ where $X \sim
\mathcal{N}(\mu=294, \sigma=10)$ and $Y\sim \mathcal{N}(\mu=1, \sigma=0.06)$, 
which is fewer rounds
than the 839 repetitions run in our evaluations under the same parameters.
The inferred parameters are then used as the prior for the response to each true
output on the x-axis (100 outputs in total): each non-faulty output is the x-axis value
multiplied by a sample from the error distribution. For the Byzantine attack scenario, 
an optimal attack of $f$ outputs are selected instead.
The blue and red shaded regions indicate the reported
68-95-99.7\% PIs\footnote{The 68\%-95\%-99.7\% PI is one, two, and three
posterior standard deviations, centered on the PC output.} for each PC output,
and the green region indicates the convex hull at each VC output.  
The presented PIs are representations of the IG of one, two, and three
standard deviations from the PC output.
These results
demonstrate that PC outputs and their PIs capture the true output even when the
independent probability of the output is low.

%% file: related.tex
\color{black}
\section{Related Work}
\label{sec:related}

The Byzantine Generals Problem \cite{bftgenerals} was first to influence a line of work that proved
a distributed system can tolerate up to $\frac{n-1}{3}$ faulty nodes in a complete network,
where each pair of nodes are directly connected. 
This led to the development of traditional agreement protocols, known as \emph{exact agreement},
which offered efficient algorithms to tolerate Byzantine faults (e.g., PBFT \cite{pbft})
and crash faults (e.g., Paxos \cite{paxos}). 
Although many optimizations to agreement protocols~\cite{zyzzyva,hotstuff,bft-smart}
have been proposed to decrease consensus overhead and increase throughput, 
exact agreement protocols are only applicable in domains in which consistency is required, 
and applications in which approximate outputs are sufficient end up ``overpaying'' for fault-tolerance \cite{overpay-fault-tolerance, byz-exspensive}.

The problem definition of \emph{approximate agreement}, which proximal consensus provides,
was first defined in The Weak Byzantine Generals Problem \cite{weakbyzgenerals}
but the first practical solution was presented
by Dolev \emph{et.~al.} \cite{DolevLPSSW86}. This work showed that approximate agreement
on scalar values must satisfy two conditions:
(1) all non-faulty replicas must eventually halt with outputs that are within $\epsilon$ of each other,
(2) the output of each non-faulty replica must be in the range of the initial values of the non-faulty replicas. 
When outputs are vectors, as presented in \cite{byz-vector-consensus}, condition (2) requires that the outputs
of non-faulty replicas be constrained by the convex hull of the initial vector outputs of non-faulty replicas. 
Unfortunately, the area of the convex hull, including in the asynchronous case~\cite{bvc-async}, or scalar range increases as non-faulty replicas diverge due to noise. 
The limitations of current approximate agreement and convex consensus solutions is two-fold:
(1) the accuracy of the approximated output is not maximized (as shown in our evaluations), (2)
there is no way to quantify the uncertainty of the approximated output.
Rather, proximal consensus not only improves the accuracy of the approximated output but it also
quantifies the uncertainty by computing the most likely output and by providing an IG on
each consensus output.


Research in 
asymptotic agreement~\cite{LeBlancHKX11} enable replicated processes
to eventually 
ensure values approach some limit. The primary mechanism for
tolerating Byzantine replicas is to remove the $f$
largest and $f$ smallest outputs, where $n>2f$.  For $n=2f+1$ this
mechanism is equivalent to taking the median of the received values,
which is similar to the Tverberg point in one dimension~\cite{tverberg-median} in our simulated
vector consensus approach because we made no timing assumptions.
Similarly, 
In-ConcReTeS~\cite{in-concretes} requires clients to define 
availability and freshness requirements and computes the median
output if replicas provide inconsistent outputs. 
In Figures~\ref{fig:no-attack}
and~\ref{fig:byz-attack}, we demonstrate the median selected can be manipulated
and is not necessarily the most accurate approach.

Optimistic BFT \cite{optimistic-byz,chai-byz} is a BFT event-streaming solution that operates with continuous one-way messaging as long as every computation round
results in a quorum of $2f+1$ matching outputs. 
Unfortunately, even if an approximate output is available, inconsistent outputs
will cause the system to stall until the inconsistency is resolved.
Conversely, the consumers in proximal consensus can define an acceptable interval width
in which a proximal consensus output based on inconsistent outputs can be accepted.
Similarly, ASPAS \cite{aspas} leverages two concurrent phases: frontend and backend.
The frontend phase lets clients query directly from one appserver, and the
backend phase enables appservers to coordinate their operation log via a BFT protocol. 
However, clients may have to rollback if the appserver in which they
queried is Byzantine, whereas the influence of Byzantine replicas is always bounded 
by the least likely $f$ non-faulty replicas or the AIW the client specifies. 

Igor~\cite{igor} executes each execution path in parallel when replicas produce
inconsistent outputs until the correct path is determined, at which point the 
invalid paths are terminated. 
Instead, the hybrid configuration of proximal consensus
allows the system to continue processing while bounding Byzantine influence in the one-shot variant, 
where divergence between non-faulty replicas is eventually corrected
by replicas also running coordinated PC. 

Approximate agreement in consensus resilient control \cite{consensual-resilient} requires the voter 
to identify when values are sufficiently close for the outputs to be accepted. However, 
since the paper discusses the difficulty of doing so, it proposes to alternatively have replicas 
first reach agreement on the sensed values or to agree on the state update before presenting 
the outputs to the voter. We believe proximal consensus would be useful here.

%% file: conclusion.tex
\section{Conclusion}
\label{sec:conclusion}

Proximal consensus (PC) computes the statistically most 
likely output conditioned on the similarity relationship between
a set of outputs while also bounding the influence of Byzantine attackers to within the an interval
proportional to the variance of non-faulty outputs.
In this paper, our simulations show a significant increase in accuracy compared to 
a Tverberg-based vector consensus (VC) protocol, we give analytical bounds on the maximal
impact by Byzantine attackers, and compare the interval guarantee and convex hull 
guarantee of PC and VC. Finally, we provide asymptotic run-time complexities and show
PC is significantly more efficient than convex hull-based protocols in higher dimensions.

%% file: appendix.tex
\onecolumn
\section* {Appendix}

\begin{algorithm}
    \caption{Proximal consensus for univariate Gaussian distribution with unknown parameters}
	\begin{algorithmic}[1]
	    \State $\mu_0$, $\nu$, $\alpha$, $\beta$ \Comment{Conjugate prior parameters of Gaussian distribution with unknown parameters~\cite{nig-conjugate}}	    
		\For { Each every quorum $q \in [Q]^{2f+1}$ } \Comment{Combinations of size $2f+1$}
		    \State $\mu', \nu, \alpha, \beta \gets$ Update conjugate prior based on $q, \mu_0, \nu, \alpha, \beta$ \Comment{Equations \ref{eq-conjugate-prior}-\ref{eq-sample-mean} in Appendix}
            \State $\sigma'^2 \gets \frac{\beta}{\alpha-1}$ \Comment{Compute the mean of the distribution variance represented by an inverse-gamma distribution~\cite{nig-conjugate}}
		  \For { Each $x_i \in X$ where $X=[\mu'-3\sqrt{\sigma'^2}, \mu'+3\sqrt{\sigma'^2}]$ } \Comment{$X$ is a discretized search space}
		        \State $C_{(x_i,q, IG)} \gets$ Calculate conditional probability $P(x_i~|~q)$ \Comment{Definition \ref{def-conditional-probability}}
		    \EndFor
		\EndFor
		\State\Return $(x_i, q), IG \in C$ \textsf{s.t.} $\max(C)$ \Comment{Return $(x_i,q)$ most conditionally likely and the interval guarantee}
	\end{algorithmic} 
	\label{alg:prox-consensus}
\end{algorithm}

\begin{algorithm}
    \caption{Replica $r$---One-shot Proximal Consensus}
	\begin{algorithmic}[1]
            \State $\mu_0$, $\nu$, $\alpha$, $\beta$ \Comment{Conjugate prior parameters of Gaussian distribution with unknown parameters~\cite{nig-conjugate}}	    
            \While {$|Q| < 3f+1$ || $IG_{PC(Q,|Q|,f)} < AIW$} \Comment{See Section~\ref{sec:system-model}}
                \State $Q \gets$ wait for additional inputs \Comment{$Q$ is the received output set}
            \EndWhile    
            \State $(x_r,q), IG \gets PC(Q,|Q|,f)$ \Comment{See Algorithm~\ref{alg:prox-consensus}}
            \State Output $x_r, IG$ \Comment{Output $x_r$ as the next stream output and the interval guarantee} 
            \State $\mu', \nu, \alpha, \beta \gets$ Update conjugate prior based on $q, \mu_0, \nu, \alpha, \beta$ \Comment{Equations \ref{eq-conjugate-prior}-\ref{eq-sample-mean} in Appendix}
	\end{algorithmic} 
	\label{alg:one-shot-prox}
\end{algorithm}

\begin{algorithm}
    \caption{Replica $r$---Coordinated Proximal Consensus}
	\begin{algorithmic}[1]
	    \State $\mu_0$, $\nu$, $\alpha$, $\beta$ \Comment{Conjugate prior parameters of Gaussian distribution with unknown parameters~\cite{nig-conjugate}}            
            \State $Q' \gets$ Run BA on $Q$ \Comment{Byzantine agreement algorithm on all proposed outputs $Q$}            
            \State $(x_r, q), IG \gets PC(Q',|Q'|,f)$ \Comment{See Algorithm~\ref{alg:prox-consensus}}
            \State Broadcast $x_r$ to client \Comment{Client is guaranteed to receive $n-f$ matching $x_r$'s}   
            \State $\mu', \nu', \alpha', \beta' \gets$ Update conjugate prior based on $q, \mu_0, \nu, \alpha, \beta$ \Comment{Equations \ref{eq-conjugate-prior}-\ref{eq-sample-mean} in Appendix}
            \If {$t > t'$} \Comment{$t$ is current time, $t'$ is next parameter update time}
                \State $(\hat{\mu}, \hat{\nu}, \hat{\alpha}, \hat{\beta})_t \gets$ Run BA on $\mu'$, $\nu'$, $\alpha'$, $\beta'$ \Comment{Byzantine agreement on posterior parameters for checkpoint $t$}
                \State Broadcast $(\hat{\mu}, \hat{\nu}, \hat{\alpha}, \hat{\beta})_t$  \Comment{Send updated parameters to one-shot replicas}
            \EndIf
	\end{algorithmic} 
	\label{alg:coordinated-prox}
\end{algorithm}

The univariate Gaussian conjugate update function~\cite{nig-conjugate} with unknown mean and variance
expressed by the prior parameters of a normal-inverse gamma distribution $\mu_0, \nu, \alpha, \beta$, sample mean $\bar{x}$ (Equation~\ref{eq-sample-mean}), and number of observations $n$:
\begin{align}    
    \mu' &= \frac{\nu\mu_0 + n\bar{x}}{\nu+n} \label{eq-conjugate-prior} \\
    \nu' &= \nu + n \\
    \alpha' &= \alpha + \frac{n}{2} \\
    \beta' &= \beta + \frac{1}{2}\sum_{i=1}^n (x_i-\bar{x})^2 + \frac{n\nu}{\nu+n}\frac{(\bar{x} - \mu_0)^2}{2} 
\end{align}

Equation for sample mean given $N$ observations on variable $X$:
\begin{equation}
\label{eq-sample-mean}
    \bar{x} = \frac{1}{N}\sum_{i=1}^{N} X_i
\end{equation}

Mathematics derivation for computing $P(X~|~q=[h_1,h_2,h_3])$:
\begin{align}
     &P(X=x~|~h_1\cap h_2 \cap h_3)  \notag \\            
    &= P(x)^{ \zeta^{1-P(h_1\cap h_2 \cap h_3)} } \label{eq:derive1} \\
    &= P(x)^ { \zeta^{1-P(h_1|h_2\cap h_3)\cdot P(h_2|h_3)\cdot P(h_3)} } \label{eq:derive2} \\        
    &= P(x)^{\zeta^{1-P(h_1)^{\psi^{1-P(h_2\cap h_3)}}  \cdot P(h_2)^{\gamma} \cdot P(h_3)}} \\
    &= P(x)^{\zeta^{1-P(h_1)^{\psi^{1-P(h_2|h_3)\cdot P(h_3)}} \cdot P(h_2)^{\gamma} \cdot P(h_3)}} \\
    &= P(x)^{\zeta^{1-P(h_1)^{\psi^{1-P(h_2)^{\gamma} \cdot P(h_3)}} \cdot P(h_2)^{\gamma} \cdot P(h_3)}}    
\end{align}
\begin{align}
    \text{where } \gamma=\Psi^{1-P(h_3)} && \notag 
      && \zeta = \frac{1-\similar([x,h_1,h_2,h_3])}{1+\similar([x,h_1,h_2,h_3])} &\notag 
      &&\Psi = \frac{1-\similar([h_1,h_2,h_3])}{1+\similar([h_1,h_2,h_3])} \notag 
\end{align}

Process to compute the similarity of candidate output $x$ and quorum $q$.
Given $D=[x, q]$ and the dimensionality of the outputs $d$, we create a list of
coordinate points of the value and its PDF: $P=\{(p_1,\ldots, p_{d},f(p,\nu))
| p \in D\}$ value using the Student-t PDF function that relies on the $\Gamma$
function~\cite{gamma-function} and a degrees of freedom parameter $\nu$
where $\nu \rightarrow \infty$ signifies the distribution's
convergence to a Gaussian distribution and is determined based on the number of observations~\cite{student-t-cdf} where:
\begin{equation}
    \label{eq-student-t-pdf-value}
    f(x, \nu) = \frac{\Gamma(\frac{\nu+1}{2})}{\sqrt{\pi\nu}\Gamma(\frac{\nu}{2})} (1+\frac{x^2}{\nu})^{-\frac{\nu+1}{2}} 
\end{equation}
We then
use the min-max normalization function independently on each index of $P$ to normalize the values on a $[0,1]$ scale:
\begin{equation}
    \label{eq-min-max-scaling}
    P_i^\textsf{norm} = \forall_{x\in P_i} \frac{x-\min(P_i)}{\max(P_i)-\min(P_i)}
\end{equation}
where $P_i$ is the collection of all values from index $i$ from $P$.
A generalized similarity function is:
\begin{equation}
    \label{eq:similarity-function}    
    \similar(P^{\textsf{norm}}) = \frac{1}{1+ \dist(P^{\textsf{norm}})} 
\end{equation}
where $\dist(P)$ is the 
cumulative $d+1$ Euclidean distance~\cite{danielsson1980euclidean} metric between each normalized coordinate pair $P^{\textsf{norm}}$:
\begin{equation}
    \label{eq-2d-dist}
    d(P^{\textsf{norm}}) = \sqrt{\forall_{p,q \in P} \sum_{i=1}^{d+1} (p_i-q_i)^2}
\end{equation}